\documentclass[10pt,a4paper,twoside]{article}
\usepackage{epsfig}
\usepackage{baltlat6}
\usepackage{array}
\usepackage{here}
\begin{document}
\ \
\vspace{0.5mm}
\setcounter{page}{1}
\vspace{8mm}

\titlehead{Baltic Astronomy, vol.\,17, 1--6, 2011}

\titleb{Modeling of the Continuous Absorption of\\ Electromagnetic Radiation in Dense Hydrogen Plasma}

\begin{authorl}
\authorb{A.~A.~Mihajlov}{1,2} and
\authorb{N. M.~Sakan}{1} and
\authorb{V.~A.~Sre{\' c}kovi{\' c}}{1,2} and
\authorb{Y.~Vitel}{3}
\end{authorl}

\begin{addressl}
\addressb{1}{Institute of Physics, Belgrade University,\\  Pregrevica 118, Zemun, 11080
         Belgrade, Serbia}
\addressb{2}{Isaac Newton Institute of Chile, Yugoslavia Branch, \\ Volgina 7,
       11060 Belgrade Serbia}
\addressb{3}{UPMC Univ Paris 6, Laboratoire des Plasmas Denses,\\ 3 rue Galilee, 94200 Ivry sur Seine, France}
\end{addressl}

\submitb{Received: 2011 June 10; accepted: 2011 December 15}

\begin{summary} In this work is examined a new modeling way of describing the
continuous absorption of electromagnetic (EM) radiation in a dense
partially ionized hydrogen plasmas with electron densities about
$5\cdot10^{18}$ cm$^{-3}$ - $1.5\cdot10^{19}$cm$^{-3}$ and
temperatures about $1.6 \cdot10^{4}$ K - $2.5 \cdot 10^{4}$ K in the
wavelength region $300 \textrm{nm} < \lambda < 500 \textrm{nm}$.  The obtained results
can be applied to the plasmas of the partially ionized layers of
different stellar atmospheres.
\end{summary}


\resthead{The modeling of the continuous of EM absorption in dense H plasmas}
{Mihajlov et al.}

\sectionb{1}{INTRODUCTION}

In this paper testing is started of a new
model way of describing some of atomic photo-ionization processes in
dense strongly ionized plasmas, which is based on the approximation
of cut-off Coulomb potential. By now this approximation has been
used only in order to describe transport properties of dense plasmas
(see for example Mihajlov et al. (1989)), but it was clear that it could be
applied to the mentioned absorption processes in non-ideal plasmas
too. Because of exceptional simplicity of the hydrogen atom, for the
first application of the mentioned approximation the following
photo-ionization processes are chosen here:

\begin{equation}
    \label{eq:ph}
    \varepsilon _{\lambda} + H^{*}(n,l) \to H^{+} + e_{E},
\end{equation}
where $\epsilon _{\lambda}$ is the energy of the photon with
wavelength $\lambda$, $n$ and $l$ - principal and orbital quantum
numbers of hydrogen atom excited states, $e_{E}$ - the free electron
in one of the states with energy $E = \hbar^{2}k^{2}/2m$, and $m$
and $\hbar$ - the electron mass and Plank's constant. It is clear
that describing the processes of the type of Eq.~(\ref{eq:ph}) in
strongly non-ideal plasmas is one of the most complicated problems.
Namely, while in weakly and moderately non-ideal plasma the
interaction of an excited atom with its neighborhood can be
neglected, as for example in Solar photosphere (Mihalas (1978); Mihajlov et al. (2007)),
or described within the framework of a perturbation theory, this is
not possible in strongly non-ideal plasmas. This is due to the fact
that in such plasmas the energy of the mentioned interaction reaches
the order of the corresponding ionization potential.

Here a new model method having a semi-empirical character of determination of the
spectral absorbtion coefficients, characterizing the bound-free
(photo-ionization) processes (\ref{eq:ph}) in strongly non-ideal
hydrogen plasmas, is presented. As
landmarks we take hydrogen plasmas with electron densities $N_e \sim
1 \cdot 10^{19} \textrm{cm}^{-3}$ and temperatures $T \approx 2\cdot 10^{4}
\textrm{K}$, which were experimentally studied in Vitel et al. (2004). The presented
method is tested within the optical range of photon wavelengths $350 \textrm{nm}
\le \lambda_{h\nu} \le 500 \textrm{nm}$.

\sectionb{2}{THEORY}

\begin{figure}[ht]
\begin{minipage}[b]{0.5\linewidth}
\centering
\includegraphics[scale=0.3]{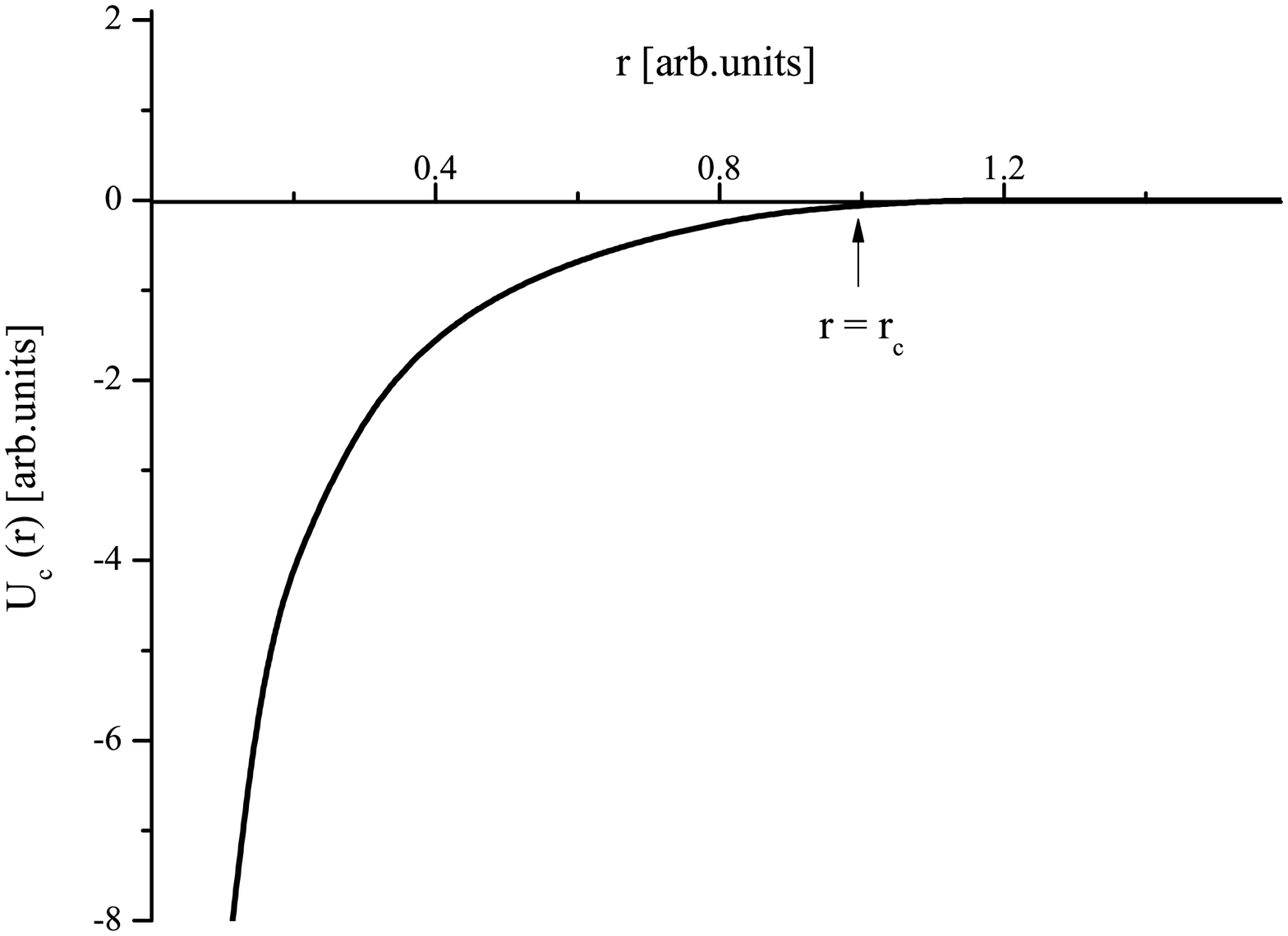}
\caption{Cut-off potential $U_{c}(r)$, where $r_{c}$ is cut-off parameter.}
\label{fig:figure1}
\end{minipage}
\hspace{0.5cm}
\begin{minipage}[b]{0.5\linewidth}
\centering
\includegraphics[scale=0.35, angle=270]{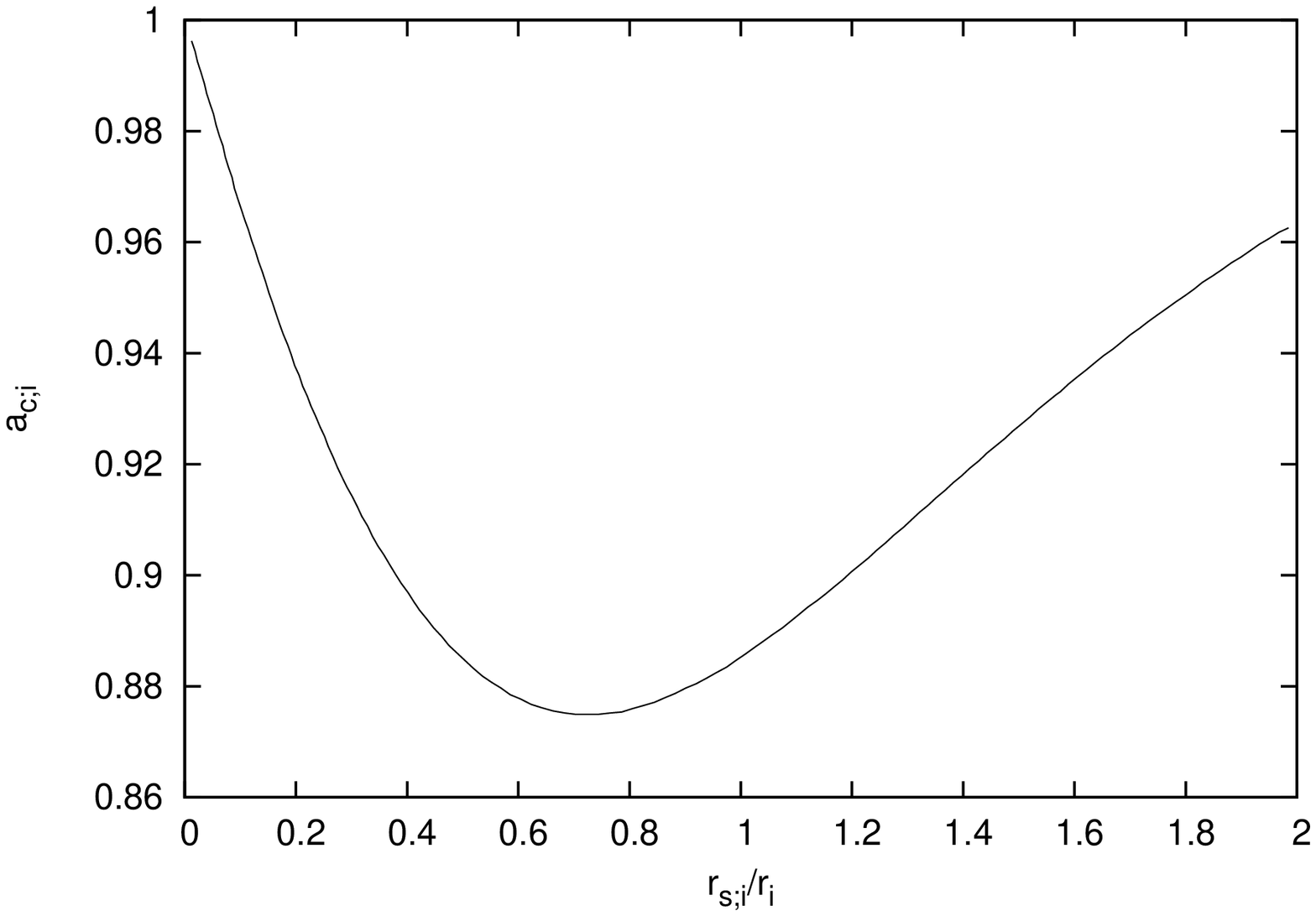}
\caption{The parameter $a_{c;i}\equiv r_{c}/r_{i}$ as the function of the
ratio $r_{s;i}/r_{i}$.}
\label{fig:figure2}
\end{minipage}
\end{figure}

The absorption processes (\ref{eq:ph}) in non-ideal
plasma are considered here as a result of radiative transition in
the whole system "electron-ion pair (atom) + the neighborhood",
namely: $\epsilon_{\lambda}+(H^{+}+e)_{n,l}+S_{rest}\to
(H^{+}+e)_{E}+S^{'}_{rest}$, where $S_{rest}$ and $S^{'}_{rest}$
denote the rest of the considered plasma. However, as it is well
known, many-body processes can sometimes be simplified by their
transformation to the corresponding single-particle processes in an
adequately chosen model potential. Here, in accordance with the
previous paper (Mihajlov et al. 1989) the screening cut-off Coulomb potential
is taken as an adequate model potential, which can be presented in the
form
\begin{equation}
\label{eq:U0} U_{c}(r) = \left\{
\begin{array}{c c}
- \displaystyle e^2/r + \displaystyle e^2/r_c, \qquad 0 < r \le
r_{c},
\\
\displaystyle 0 ,              \qquad \qquad      r_{c} < r <
\infty,
\end{array}
\right.
\end{equation}
which is illustrated by Fig.~\ref{fig:figure1}. Here $e$ is the modulus of the
electron charge, $r$ - distance from the ion, and cut-off radius
$r_{c}$ - the characteristic screening length of the considered
problem. Namely, within this model it is assumed that quantity
$U_{p;c}=-e^{2}/r_{c}$ is the mean potential energy of an electron
in the considered hydrogen plasma. It is important that the cut-off
radius $r_{c}$ can be determined as a given function of $N_{e}$ and
$T$, using two characteristic lengths: $r_{i} =[k_{B}T/(4 \pi N_{i} e^{2})]^{1/2}$
and $r_{s;i} = [3/(4\pi N_{i})]^{1/3}$, where
$N_{i}$ and $r_{s;i}$ are the $H^{+}$ density and the corresponding
Wigner-Seitz's radius and $k_{B}$ - Boltzman's constant. Namely,
taking that $N_{i}=N_{e}$ and
\begin{equation}
\label{eq:pci} r_{c} = a_{c;i}\cdot r_{i},
\end{equation}
we can directly determine the factor $a_{c;i}$ as a function of
ratio $r_{s;i}/r_{i}$, on the basis of the data about the mean
potential energy of the electron in the single ionized plasma from
Mihajlov et al. (2009). The behavior of $a_{c;i}$ in a wide region of values
of $r_{s;i}/r_{i}$ is presented in Fig.\ref{fig:figure2}.

In diluted hydrogen plasma (see for example Mihalas 1978) the
spectral absorption coefficients, characterizing the
photo-ionization processes (\ref{eq:ph}) can be described within the
approximation of the non-perturbed energy levels in the potential
$U_{c}(r)$, namely: $ \kappa _{ph} ^{(0)} (\lambda; N_{e}, T) =
    \sum _{n,l}N_{n,l} \cdot \sigma_{ph}(\lambda;n,l,E_{n,l}),
$ where $N_{n,l}$ is the density of the atoms in the realized excited
states with given $n$ and $l$, and
$\sigma_{ph}(\lambda;n,l,E_{n,l})$ - the corresponding
photo-ionization cross section for $n \ge 2$. According to the
above mentioned, it cannot model the absorption coefficients of the
dense non-ideal plasmas described in Vitel et al. (2004).

It can be shown, using the results from Adamyan (2009), that $\kappa
_{ph}(\lambda; N_{e}, T)$ can be obtained within the approximation based
on adequately chosen shifts $\Delta_{n,l}$ and broadenings
$\delta_{n,l}$ of the energy levels with given $n$ and $l$. It is
assumed that energies $\epsilon$ of the perturbed atomic states are
dominantly grouped around energy $\epsilon_{n,l}^{(max)}= E(n,l) +
\Delta_{n,l}$, inside the interval $(\epsilon_{n,l}^{(max)} -
\delta_{n,l}/2, \epsilon_{n,l}^{(max)} + \delta_{n,l}/2)$, similarly
to the known cases (Gaus, Lorentz, uniform etc.).

Let us note that it is possible to describe the quantity
$\Delta_{n}$ as a function of $N_{e}$. Namely, for well-known
physical reasons all shifts $\Delta_{n,l}$, and consequently
$\Delta_{n}$, have to change proportionally with the density of the
perturbers, the relative atom-perturber velocity and the
characteristic perturbation energy. Consequently, we will have
that: $\Delta_{n} \sim N_{e}\cdot v_{ea}(T)\cdot e^{2}/l(N_{e},T)$,
where $v_{ea}(T)$ and $l(N_{e},T)$ are the characteristic
electron-atom velocity and distance. On the basis of the results of
Mihajlov et al. (2009) in the considered cases ($N_{e}\sim 1\cdot 10^{19}
\textrm{cm}^{-3}$, $T \sim 2 \cdot 10^{4} \textrm{K}$) any relevant characteristic
length has to be close to the radius $r_{i}$. From here, since $v_{ea}(T)\sim
(k_{B}T)^{1/2}$ and $r_{i}\sim(k_{B}T/N_{e})^{1/2}$, the relation
follows
\begin{equation}
\label{eq:Deltan} \Delta_{n} \approx Const. \cdot N_{e}^{3/2},
\end{equation}
which is in accordance with Adamyan (2009) and can be useful in
further considerations.

Here, we will describe the perturbed atomic states in the first
order of the perturbation theory and, in accordance with what was said
above, we will have it that
\begin{equation}
\label{eq:kappaph} \kappa_{ph}(\lambda; N_{e}, T) =  \sum
_{n,l}N_{n,l} \cdot \frac{1}{\delta_{n}} \int
\limits_{\epsilon_{n,l}^{(max)}-\delta_{n}/2}^ {\epsilon_{n,l}^{max}
+ \delta_{n}/2} {\frac{\varepsilon_{\lambda}}{\varepsilon_{\lambda}
+ \epsilon} \cdot\sigma_{ph}(\lambda^{(\epsilon)}; n, l, E_{n,l})}
d\epsilon,
\end{equation}
where $n \ge 2$, $\epsilon_{n,l}^{(max)} = E_{n,l} + \Delta_{n}$,
and $\sigma_{ph}(\lambda^{(\epsilon)}; n,l,E_{n,l})$ is the corresponding
photo-ionization cross section for $\lambda^{(\epsilon)}= \lambda \cdot
\varepsilon_{\lambda}/ (\varepsilon_{\lambda} + \epsilon)$, i.e. for
the wavelength of the photon with energy $(\varepsilon_{\lambda}
+ \epsilon)$.

\sectionb{3}{RESULTS AND DISCUSSION}

In this paper the approximation of
cut-off Coulomb potential (\ref{eq:U0}) is applied to the modeling
of spectral absorption coefficients of the hydrogen plasma, obtained
in Vitel et al. (2004) in two experiments: a short and a long pulse,
respectively. In the first case (short pulse) plasma with $N_e= 1.5
\cdot 10^{19} \textrm{cm}^{-3}$ and $T = 2.3\cdot 10^{4} \textrm{K}$ was studied,
while in the second case (long pulse) - one with $N_{e}= 6.5 \cdot
10^{18} \textrm{cm}^{-3}$ and $T =1.8\cdot 10^{4} \textrm{K}$.  It has been found
that: $r_{c} = 44.964$ a.u. for a short pulse, and $r_{c} = 55.052$
a.u. for the long one.

In order to compare the obtained theoretical results with the
experimental data from Vitel et al. (2004), we had to take into account
other relevant absorption processes, namely: $(e + H^{+})$-inverse
"bremsstrahlung", as well as $H^{-}$ and $H_{2}^{+}$ absorption
continuums, which cannot be neglected in the considered hydrogen
plasmas. Therefore, when comparing our theoretical results with the
experimental data from Vitel et al. (2004) we use the total spectral
absorption coefficient $\kappa_{tot}(\lambda)$ given by:
$\kappa_{tot}(\lambda) = \kappa_{ph}(\lambda) +
\kappa_{add}(\lambda)$, where the member $\kappa_{ph}(\lambda)\equiv
\kappa_{ph}(\lambda;N_{e},T)$ is given by Eq.~(\ref{eq:kappaph}),
while the member $\kappa_{add}(\lambda)$ is the sum of the
absorption coefficients of all additional processes. Let us note
that electron-ion process is described by the absorption coefficient
from Sobel'man (1979), while the electron-atom and ion-atom processes -
by the ones determined as in previous paper (Mihajlov et al. 2007) dedicated
to the same absorption processes in the solar photosphere.

\begin{figure}[ht]
\begin{minipage}[b]{0.5\linewidth}
\centering
\includegraphics[width=\columnwidth,
height=0.75\columnwidth]{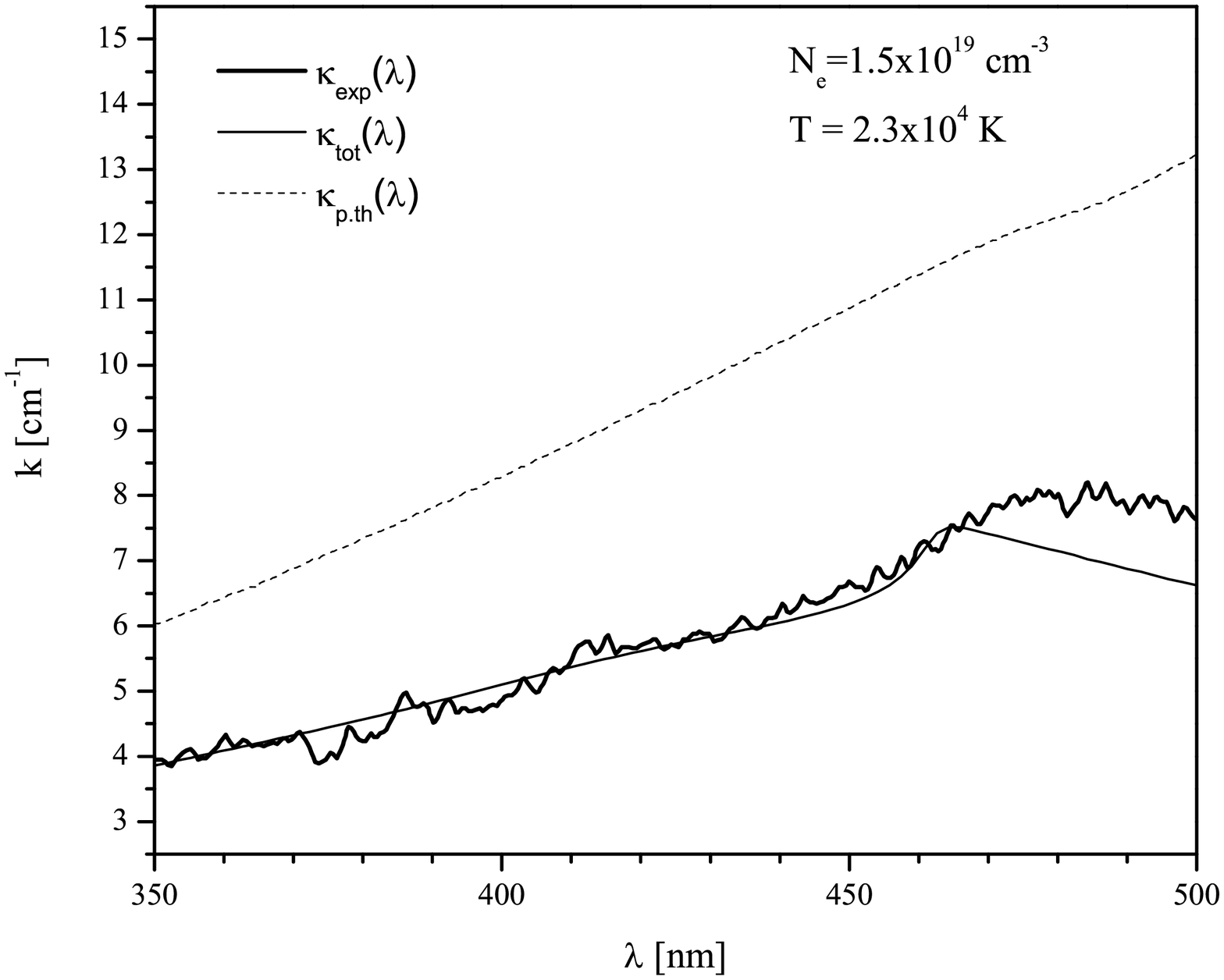}
\caption{The absorbtion coefficient $\kappa_{tot}(\lambda)$ calculated in
the case of the short pulse with $\Delta_{n}=0.455\textrm{eV}$ and
$\delta_{n}=0.625\textrm{eV}$. Dashed line - the theoretical curve from
Vitel et al. (2004).}
\label{fig:figure3}
\end{minipage}
\hspace{0.5cm}
\begin{minipage}[b]{0.5\linewidth}
\centering
\includegraphics[width=\columnwidth,
height=0.75\columnwidth]{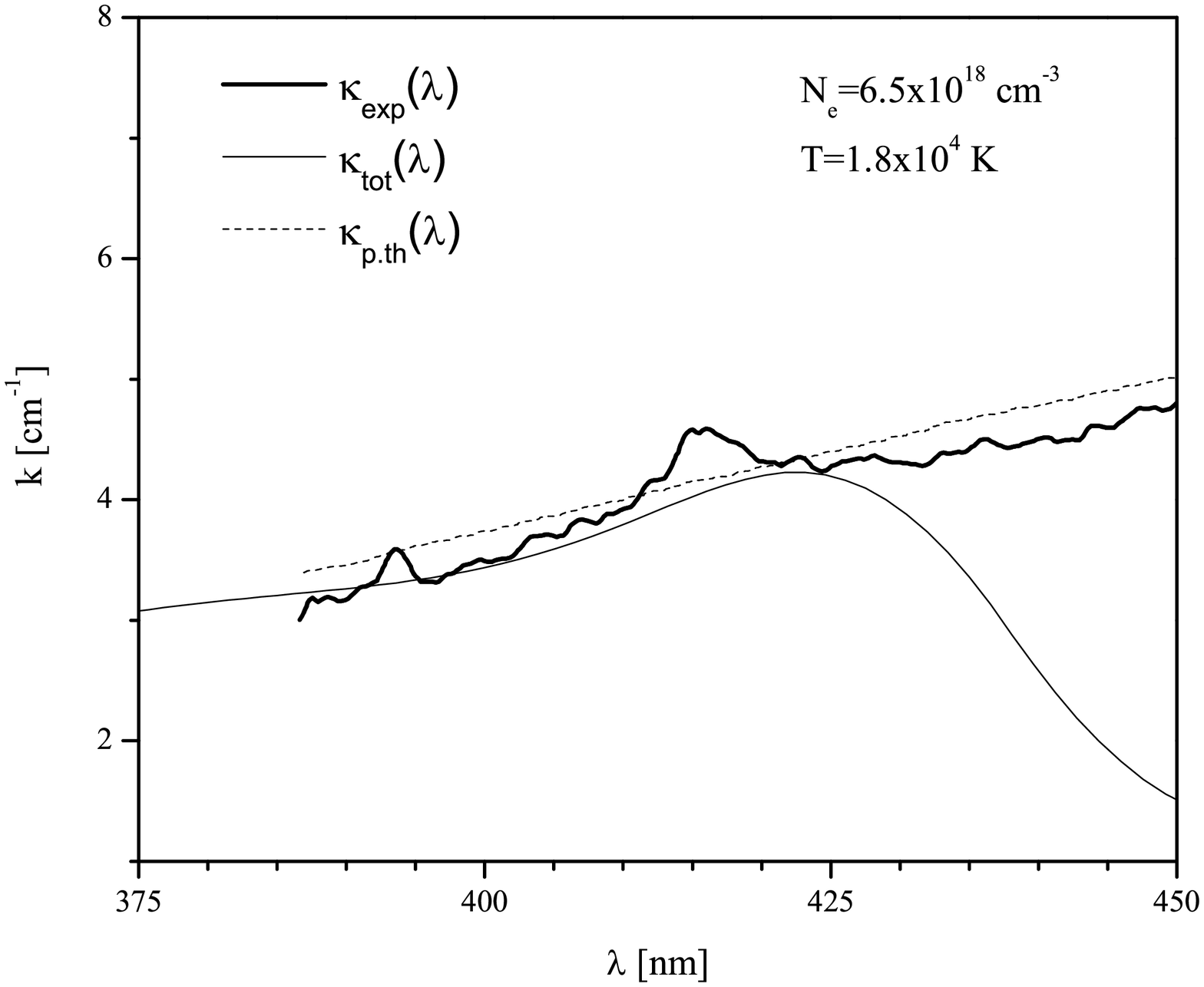}
\caption{The absorbtion coefficient $\kappa_{tot}(\lambda)$ calculated in
the case of the Long pulse with $\Delta_{n}=0.13\textrm{eV}$ and
$\delta_{n}=0.11\textrm{eV}$. Dashed line - the theoretical curve from
Vitel et al. (2004).}
\label{fig:figure4}
\end{minipage}
\end{figure}

In accordance with the aims of this work the calculations of the
total absorbtion coefficient, are performed for both cases
(short and long pulse) in wide regions of
values of shifts ($\Delta_{n}$) and broadening ($\delta_{n}$) of
atomic levels with $n \ge 2$. The calculations of
$\kappa_{tot}(\lambda)$ cover wavelength region $350\textrm{nm} \le \lambda
\le 500\textrm{nm}$. The results of calculations are shown in Figs.~\ref{fig:figure3}
and \ref{fig:figure4} together with the corresponding experimental values
$\kappa_{exp}(\lambda)$ of the spectral absorbtion coefficient from
Vitel et al. (2004). This figures show the results of the calculations of
$\kappa_{tot}(\lambda)$ in the case $\Delta_{n} = const.$, with the
values of $\Delta_{n}$ and $\delta_{n}$ which are treated as optimal
ones: $\Delta_{n}= 0.455 \textrm{eV}$ and $\delta_{n}= 0.625 \textrm{eV}$ for the
short pulse, and $\Delta_{n}=0.13 \textrm{eV}$ and $\delta_{n}= 0.11 \textrm{eV}$ for
the long pulse.

Here it is important to check whether relation Eq.~(\ref{eq:Deltan})
is valid also for $N_{e}$ close to $0.65\cdot 10^{19} \textrm{cm}^{-3}$.
Since in the case of constant shifts $\Delta_{n}=0.455 \textrm{eV}$ and
$0.130 \textrm{eV}$ for short and long pulses respectively, validity of
Eq.~(\ref{eq:Deltan}) means that $0.455/0.130=(1.5/0.65)^{3/2}$,
which is satisfied with an accuracy better than $1\%$. In the case
of variable shift we have it that $\Delta_{n=2}=0.49 \textrm{eV}$ and $0.12
eV$ for the short and long pulses respectively, and validity of
Eq.~(\ref{eq:Deltan}) means now that $0.49/0.14=(1.5/0.65)^{3/2}$,
which is satisfied with the same accuracy. The fact that
Eq.~(\ref{eq:Deltan}) is satisfied for $N_{e}=1.5\cdot 10^{19}
\textrm{cm}^{-3}$ and $0.65\cdot 10^{19} \textrm{cm}^{-3}$ offers a possibility to
determine $\Delta_{n}$ or $\Delta_{n=2}$ not only for these
densities but also for any $N_{e}$ from interval $0.65\cdot 10^{19}
\textrm{cm}^{-3} < N_{e} < 1.5\cdot 10^{19} \textrm{cm}^{-3}$ and probably in a
significantly wider region. Also, using the fact that the influence
of $\delta_{n}$ over the absorbtion coefficients is significantly
weaker than the influence of $\Delta_{n}$, we can determine the
values of $\delta_{n}$ for any $N_{e}$ using the values of ratio
$\delta_{n}/\Delta_{n}$ from the considered examples.

On the grounds of all that was said one can conclude that the
presented method can already be used for calculations of the
spectral absorbtion coefficients of dense hydrogen plasmas with
$N_{e}\sim 10^{19} \textrm{cm}^{-3}$ and $T_{e}\approx 2 \cdot 10^{4}\textrm{K}$. Let
us note that, with some minor modifications, the presented method
can be applied to any kind of dense single-ionized plasma
(laboratorial alkali-metal plasmas, helium plasmas in some DB white
dwarfs etc.

\thanks{The authors are thankful to the University
P. et M. Curie of Paris (France) for financial support, as well as
to the Ministry of Science of the Republic of Serbia for support
within the Projects 176002, III44002 and 171014.}

\References

\refb Adamyan V. M. 2009, {\it private communication}

\refb Mihajlov A. A.,Djordjevic D., Popovic M. M. et al. 1989, Contrib. Plasma Phys., 29, 441

\refb Mihajlov A. A., Ignjatovi\'c L. M., Sakan N. M. et al. 2007, A\&A, 437, 1023

\refb Mihajlov A. A., Vitel Y., Ignjatovic L. M. 2009, High Temperature, 47, 5

\refb Mihalas D. 1978, in {\it Stellar Atmospheres},
San Francisco

\refb Sobel'man I. I. 1979, in {\it Atomic Spectra and Radiative Transitions},
Springer Verlag, Berlin

\refb Vitel Y., Gavrilova T. V., D'yachkov L. G. et al. 2004, JQSRT, 83, 387

\end{document}